\PassOptionsToPackage{english}{babel}
\documentclass[useAMS,usenatbib]{mn2e}

\usepackage[english]{babel}
\usepackage{amsmath}
\usepackage{color}
\usepackage{graphicx}
\usepackage{breqn}
\usepackage{amssymb}
\usepackage{tabu}
\usepackage[draft]{hyperref}
\usepackage{caption}
\usepackage{booktabs,caption}
\usepackage{subcaption}
\usepackage{hyperref}
\hypersetup{
    pdfproducer={}
}

\title[Observability of CPDs]{Observability of Forming Planets and their Circumplanetary Disks III. -- Polarized Scattered Light in Near-IR}
\author[Szul\'agyi \& Garufi]{
\parbox{\textwidth}{
J. Szul\'agyi$^{1,2}$\thanks{E-mail:
judit.szulagyi@phys.ethz.ch} \& A. Garufi$^{3}$,\\
}\vspace{3mm}\\
$^{1}$ Institute for Particle Physics and Astrophysics, ETH Z\"urich, Switzerland\\
$^{2}$ Center for Theoretical Astrophysics and Cosmology, Institute for Computational Science, University of Z\"urich, Switzerland\\
$^{3}$ INAF, Osservatorio Astrofisico di Arcetri, Firenze, Italy\\
}

\begin{document}

\date{Accepted XX. Received XX; in original form 2019 May 17}

\pagerange{\pageref{firstpage}--\pageref{lastpage}} \pubyear{2019}

\maketitle

\label{firstpage}

\begin{abstract}
There are growing amount of very high-resolution polarized scattered light images of circumstellar disks. Nascent giant planets planets are surrounded by their own circumplanetary disks which may scatter and polarize both the planetary and stellar light. Here we investigate whether we could detect circumplanetary disks with the same technique and what can we learn from such detections. Here we created scattered light mock observations { at 1.245 microns (J band) for instruments like SPHERE and GPI, for various planetary masses (0.3, 1.0, 5.0, 10.0 $\rm{M_{Jup}}$), disk inclinations (90, 60, 30, 0 degrees) and planet position angles (0, 45, 90 degrees).} We found that the detection of a circumplanetary disk at 50AU from the star is significantly favored if the planet is massive ($\geq 5 \rm{M_{Jup}}$) and the system is nearly face-on ($\leq 30^\circ$). {In these cases the accretion shock front on the surface of the circumplanetary disks are strong and bright enough to help the visibility of this subdisk.} Its detection is hindered by the neighboring circumstellar disk that also provides a strong polarized flux. However, the comparison between the $PI$ and the $Q_\phi$ maps is a viable tool to pinpoint the presence of the circumplanetary disk within the circumstellar disk, as the two disks are behaving differently on those images.
\end{abstract}


\section{Introduction}

Young, forming giant planets are surrounded by their circumplanetary disks (CPDs), where their regular satellites will form eventually. Regardless whether the planet formed via core accretion or disk instability scenario, the circumplanetary disk forms in the last phase of the planet formation \citep[e.g.][]{Kley99,Lubow99,SB13,AB09a,AB09b}. 

The circumplanetary disks characteristics have been studied numerically since two decades now \citep[e.g.][]{Kley99,Lubow99,DAngelo03,AB09a,AB09b,Machida10,AB12,Tanigawa12,SB13,Gressel13,Szulagyi14,Fujii14,Perez15,DAngelo15,Tanigawa14,Zhu16}. As circumstellar disk, they widely vary in mass from $10^{-4}$ $\rm{M_{planet}}$ (\citealt{DAngelo03}; G. D'Angelo private communication; \citealt{Gressel13}) till $\sim$ $\rm{M_{planet}}$ \citep{SB13,Szulagyi17GI} if the planet formed via gravitational instability. Because the CPD is constantly fed from the circumstellar disk \citep[e.g.][]{Szulagyi14,FC16}, the mass of the CPD depends on the circumstellar disk mass (apart from the planetary mass \citealt{Szulagyi17gap}). The CPD's temperature will depend on the semi-major axis of the planet, the local density and opacity, the mass of the planet and its age, the viscosity of the gas, among other factors \citep{DAngelo03,PN05,AB09b,AB12,Gressel13,Szulagyi17gap}. The temperature of the CPD vary from thousands of Kelvins in the forming planet vicinity to a few hundred Kelvins in the outer subdisk. Of course, the CPD evolves in time, similarly to the circumstellar disk, getting lighter and cooler during its lifetime \citep{Szulagyi17gap}. 

The characteristics of the circumplanetary disks are affecting their detectability, hence creating mock observations from the hydrodynamical simulations is a useful tool to plan and interpret real observations. Planet-disk interactions, such as gaps has been studied on synthetic images \citep{Dipierro15, Szulagyi17alma, DSHARP}. Circumplanetary disks had been predicted to be seen with ALMA and VLA \citep{Szulagyi17alma,IsellaTurner18,Zhu18}. Mock images of polarized light about circumstellar disks helped us understanding what polarized light observations can reveal about the circumstellar disk characteristics  \citep{Dong12}. Synthetic observations of scattered light shed light on how planet-disk interactions -- especially spirals -- are expected to look like \citep{Dong15,Dong15b,FD15,Dong16}. It has also been suggested, that polarized light from the circumplanetary disk dust could be detected in favorable circumstances \citep{Stolker}. The detectability of the CPD in near-infrared as well as via spectral energy distributions was investigated in \citet{Szulagyi19}. Actual detections of circumplanetary disks are just began in 2019: two possible candidates from ALMA continuum emission in the PDS70 disk \citep{Isella19}, as well as infrared excess detection of circumplanetary material around PDS70b \citep{Christiaens19}. With these few possible detections, the characterization of circumplanetary disks from observations are still not yet possible.

Unlike circumplanetary disks, circumstellar disks have been thoroughly characterized from observations during the last decade thanks to optical/near-IR instruments like VLT/SPHERE and GPI \citep[e.g.,][]{Garufi2017b, Rapson2015} and to the (sub-)mm interferometer ALMA \citep[e.g.,][]{Andrews2018}. Among the near-IR observations, the most successful technique to directly image circumstellar disks is currently the polarized differential imaging \citep[PDI,][]{Kuhn2001, Apai2004}. This technique allows a very good removal of the strong stellar flux by separating the polarized light (mostly scattered light from the disk) from the unpolarized light (mainly stellar light). Therefore, most of the available high-resolution near-IR maps of circumstellar disks trace the polarized scattered light from the disk surface. In principle, these polarized scattered light observations also open the way to detect the circumplanetary disk the same way although this is yet to be proven observationally.


In this paper we combine temperature-included (i.e. radiative) 3D gas hydrodynamic simulations, with Monte-Carlo radiative transfer to create mock observations about detecting the circumplanetary disk in scattered light with and without polarization. In the first paper of this series, we looked at the circumplanetary disk observability in sub-mm/radio wavelength \citep{Szulagyi17alma}. In the second paper, we reviewed the case for near-infrared and spectral energy distributions \citep{Szulagyi19}. In \citet{SzE20} we made predictions of hydrogen recombination line fluxes with extinction and determined the planet-mass/planet accretion versus H-alpha, Paschen-beta, Brackett-gamma line luminosity relationships.

\section[]{Methods}
\label{sec:numerical}

We had a three step process for creating the mock images presented in this work. First, we run 3D radiative hydrodynamic simulations of the circumstellar disk with a forming planet embedded within (Sect. \ref{sec:hydro}). Then we used the RADMC-3D radiative transfer tool to create wavelength-dependent images of the systems {on 1.245 microns} with polarization (Sect. \ref{sec:radmc3d}). Finally, we convolved the images with a diffraction limited PSF for the VLT/SPHERE instruments and created polarization maps (Sect. \ref{sec:convol}).

\subsection{Hydrodynamic Simulations}
\label{sec:hydro}

The hydrodynamic simulations in this study are the same as in our previous paper \citep{Szulagyi19} of the series. In brief, we had a circumstellar disk with a mass of $\sim 10^{-2} \mathrm{M_{Sun}}$ between 20 and 120 AU around a solar-mass star, where a planet is forming at 50 AU. In four different simulations, the planet masses were chosen to be a Saturn-mass, 1 Jupiter-mass, 5 Jupiter-masses and 10 Jupiter-masses (i.e. only one planet present in each hydrodynamic run). We used the JUPITER code to carry our the hydrodynamic calculations, that was developed by F. Masset and J. Szul\'agyi \citep{Szulagyi14,Szulagyi16a} that not only solves Euler equations but also the radiative transfer with the flux-limited diffusion approximation (two-temperature approach \citealt[e.g.][]{Kley89,Commercon11}). {The heating processes include adiabatic compression, viscous heating, shock heating and stellar irradiation, while the cooling processes are adiabatic expansion and radiative diffusion. The main source of heating in the circumplanetary disk is the accretion process \citep{Szulagyi16a}, as the gas tries to fall onto the planet, leading to adiabatic compression in this region that heats up the compressing gas. Furthermore, the accretion shock front on the circumplanetary disk surface \citep{SzM17} in case of the higher mass planets are also strongly heated up. Viscous heating in the CPD is secondary, with the chosen low viscosity value. Stellar irradiation is computed based on solar flux, but its effect on the CPD is negligible. } The viscosity was a constant kinematic viscosity of $10^{-5} \mathrm{a_{p}}^2\Omega_p$, where $ \mathrm{a_{p}}$ is the semi-major axis and $\Omega_p$ denotes the orbital frequency of the planet.

Given that we were particularly interested in the circumplanetary region, where high-resolution is necessary to get the disk characteristics (density, temperature, velocities) right, we used mesh refinement in this region. This meant that while the circumstellar disk has been simulated with a lower resolution (680 cells azimuthally over $2\pi$, 215 cells radially between 20 and 120 AU and 20 cells in the co-latitude direction over 7.4 degrees opening angle from the midplane), the Hill-sphere of the planet were well resolved with four levels of refinement. Each level doubled the resolution in each spatial direction, hence the final resolution in the planet vicinity was 0.029 AU. 

While the dust was not explicitly simulated within the hydrodynamics, its effect on the temperature of the disk is taken into account through the dust opacities (with the limit of assuming a constant dust-to-gas ratio of 1\%). The opacity table was equivalent to what was used in \citet{Szulagyi19}, and included both gas and dust opacities.

\subsection{RADMC-3D post-processing}
\label{sec:radmc3d}

RADMC-3D \citep{Dullemond12}\footnote{\url{http://www.ita.uni-heidelberg.de/~dullemond/software/radmc-3d/}} radiative transfer tool was used to create wavelength-dependent intensity images from the hydrodynamic simulations. {We used $5\times10^7$ photons for these Monte-Carlo runs. We run the RADMC-3D with the flux conservation option, which makes sure that the total flux of the images are conserved as well, regardless the image resolution. We ran two sets of images:
\begin{itemize}
    \item 5000x5000 pixel resolution image on the entire circumstellar disk; ran multiple times with planet positions of 0 deg, 45 deg, 90 deg, and inclinations of 0 deg, 30 deg, 60 deg and 90 deg. 
    \item 1000x1000 pixel images on the Hill-sphere (using \url{zoomau} command); ran 20 times with randomly changing seed number and averaged at the end. We calculated the variance between these 20 runs and verified that it is near zero, so convergence was reached. These CPD region images were produced for the four different inclinations (0 deg, 30 deg, 60 deg and 90 deg).
\end{itemize} }

The dust-density files were created from the gas density (i.e. assuming that these micron-sized dust grains are strongly coupled to the gas), by multiplying the gas density in each cell with the dust-to-gas ratio of 1\%. We assumed thermal equilibrium, hence we used the dust temperature to be equal to the gas temperature, except that the dust evaporation above 1500 K was taken into account. This meant that in the cells hotter than this limit, the dust density was set to zero, to be consistent with the radiative hydrodynamic simulation, where the opacity table contains the dust evaporation as well. {We used the hydrodynamic simulation calculated temperature for the dust, because that includes shock-heating, accretional heating, viscous heating which are very important in the circumplanetary disk region and result in a hot planet vicinity. We compared our results based on another method, where we used thermal Monte Carlo simulation to calculate the dust temperature, using \url{radmc3d mctherm}, but this method does not account for the main heating mechanisms that take place in the planet vicinity, hence result in different mock images (see discussion in Sect. \ref{sec:discussion})}.

{ The distance of the circumstellar disk was assumed to be 100 parsec for all the calculations.}

The hydrodynamic simulations cannot handle well optically thin, low-density regions of the circumstellar disk, such as the disk atmosphere. In the hydro simulations the disk opening angle was only 7.4 degrees, but real circumstellar disks have a larger opening angle. Therefore we had to extend the circumstellar disk in the vertical direction using an extrapolation technique, before we ran the RADMC-3D calculations. The extrapolation was as follows. First, we fitted Gaussian-functions to the density field in each cell column (z-direction) separately, so that the vertical extent of the circumstellar disk was 2.5 times larger than the original hydro simulation's. Second, in this circumstellar disk atmosphere region, we kept the temperature as it is in the last (optically thin) co-latitude cell. Here the temperature is high due to stellar irradiation, much higher than in the bulk of the circumstellar disk (midplane regions). This meant that the temperature in the circumstellar disk atmosphere was constant with co-latitude.

The dust opacities were identical to what had been used in \citet{Pohl17}. It was assumed to be a mixture made of silicates (\citealt{draine2003b}), carbon (\citealt{zubko1996}), and water ice (\citealt{warren2008}) with fractional abundances of 7\%, 21\%, and 42\%, consistent with \citet{ricci2010}. The remaining 30\% was vacuum. The opacity of the mixture was determined by means of the Bruggeman mixing formula. The absorption and scattering opacities, $\kappa_{\mathrm{scat}}$ and $\kappa_{\mathrm{abs}}$, as well as the scattering matrix elements $Z_{ij}$ were calculated for spherical, compact dust grains with Mie theory considering the BHMIE code of \citet{bohren1983}. The grain sizes were between 0.01 micron and 150 micron, with a power-law index of -3.5. 

\subsection{Polarization maps}
\label{sec:convol}

To compare our simulations to the available observations, we first convolved the images with a rotationally symmetric 2D Gaussian Point-Spread-Function, with a Full-width-half-maximum to be $1.22\cdot\lambda/D$, where $\lambda$ is the wavelength and $D$ is the mirror-size of 8.2 meters (equivalent of VLT mirror diameter). The RADMC-3D provides the set of Stokes parameters $I, Q, U, V$. The polarized intensity map $PI$ was obtained through:
\begin{equation}
PI = \sqrt{Q^2+U^2}
\label{eq:stokespi}
\end{equation}

An alternative treatment of the Stokes parameters is commonly used in observational work, that is the creation of the tangential (sometimes called radial or polar) parameters $Q_{\phi}$ and $U_{\phi}$ \citep{Canovas2015,Monnier19}. These are defined as:
\begin{equation}
\begin{split}
	Q_{\phi} &= +Q \cos\,2\phi + U \sin\,2\phi \,, \\
	U_{\phi} &= -Q \sin\,2\phi + U \cos\,2\phi \,
	\label{eq:stokesphi}
\end{split}
\end{equation}
\noindent with $\phi$ being the angle with respect to the stellar position (x$_{0}$,y$_{0}$) calculated as:
\begin{equation}
	\phi = \mathrm{arctan} \frac{x-x_0}{y-y_0}\,
	\label{eq:azimuth}
\end{equation}
\noindent By construction, $Q_{\phi}$ corresponds to $PI$ in the scenario of perfectly centro-symmetric scattering {and single scattering}, whereas $U_{\phi}$ is ideally expected to only contain noise.

\section{Results}
\label{sec:results}

The obtained $I,\ PI,\ Q_{\phi}$ and $U_{\phi}$ maps of {on the different inclinations (0, 30, 60, 90 degrees), and different planetary positions (0, 45, 90 degrees)}. The J-band images of zero inclination are shown in Fig. \ref{fig:scat12_incl0}, while the other inclinations are in the Appendix \ref{All_maps}. Fig. \ref{fig:scat12_incl0} compares the simulations of the four planetary masses considered: 0.3 $\rm{M_{Jup}}$, 1 $\rm{M_{Jup}}$, 5 $\rm{M_{Jup}}$, 10 $\rm{M_{Jup}}$. On the images, the planet (and circumplanetary disk) always lies to the East at 50 AU from the central star.

\begin{figure*}
\includegraphics[width=18cm]{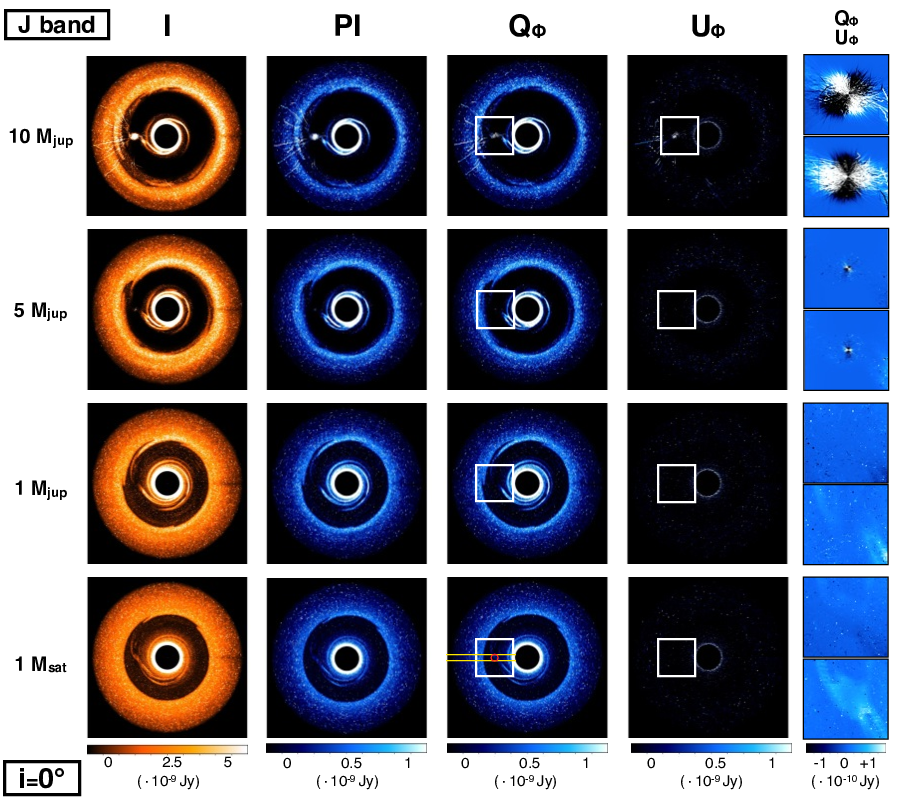}
\caption{Polarized scattered light images at 1.245 microns (J band) with 0$^\circ$ inclination for the 10, 5, 1, and 0.3 Jupiter-mass cases (from top to bottom). The columns are the $I$, $PI$, $Q_\phi$, $U_\phi$, and a zoom of the $Q_\phi$ (top) and $U_\phi$ maps (bottom) on the CPD region, respectively. The $PI$, $Q_\phi$, and $U_\phi$ images have the same color stretch to highlight their relative flux brightness, whereas the zoomed maps have a harder stretch with negative values shown in black. The white box indicates the zoomed area of the last column. The yellow and red lines on the last $Q_\phi$ map highlight the region used to calculate the contrast of CSD and CPD, respectively. {The assumed distance is 100 pc.}}
\label{fig:scat12_incl0}
\end{figure*}

From these images, the main circumstellar disk (CSD) is always very bright in $PI$ and its morphology resembles that of the $I$ images. The circumplanetary disk is visible in the first two cases only, that is with a planet of 10 and 5 $\rm{M_{Jup}}$. Similar considerations apply to the $Q_{\phi}$ images and these maps look very similar to the $PI$. On the other hand, the $U_{\phi}$ images do not show any significant signal except around the circumplanetary disk in the first case. The reason for the CPD visibility only in the high-mass planet cases, because {in these cases the accretion shock front on the surface of the CPD is strong (the velocity of the incoming accretion flow is super-sonic), creating a hot, bright surface \citep{SzM17}. In the smaller mass planet cases, the accretion flow is slower and sub-sonic, hence the shock is not as strong and not as hot (see \ref{sec:discussion} for further discussion on this point). This accretion shock is created by the incoming circumstellar disk material, through the meridional circulation, a mass transfer between the circumstellar and circumplanetary disks \citep{Szulagyi14,FC16}. The accretion shock on the circumplanetary disk surface hence contributes to the observability and observational appearance of the CPD.}


What is described above for the 0 inclination case (Fig. \ref{fig:scat12_incl0}) also applies to the other images created for the other inclinations (see Appendix \ref{All_maps}). The only obvious differences are that the circumplanetary disk becomes decreasingly evident with increasing inclination, and that some signal is recovered from the $U_{\phi}$ image when the inclination is high, in agreement with the theoretical prediction by \citet{Canovas2015}.


\begin{figure}
\includegraphics[width=8.5cm]{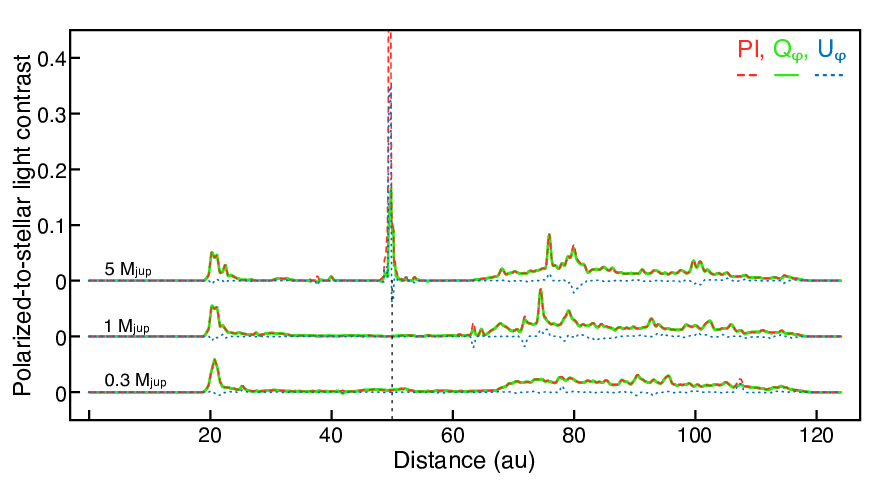}
\caption{Radial profile of the polarized-to-stellar light contrast of $PI$, $Q_{\phi}$ and $U_{\phi}$ as described in Sect.\,\ref{sec:contrast}. The profiles obtained from three different planetary masses are displaced along the y-axis for a better visualization. The vertical line indicates the planet location. {The $PI$ and $Q_{\phi}$ profiles are coincident except where the 5 $M_{\rm jup}$ mass planet is detected} (see Sect.\,\ref{sec:CSDvsCPD}).}
\label{fig:profile}
\end{figure}

\subsection{Polarized contrast}
\label{sec:contrast}

In this section, we provide a more quantitative analysis of the maps in Fig.\,\ref{fig:scat12_incl0}, as well as of those shown in Appendix \ref{All_maps}. Measuring the amount of scattered light from real observations is a challenging task because of the difficulties in flux-calibrating the images and because the disk flux is directly dependent on the stellar flux. Some authors {(e.g., \citealt{Avenhaus2018, Garufi18})} quantified the near-IR  polarized light from the disk in relation to the stellar flux, thus as to alleviate the dependence on the stellar brightness. In particular, a way to do it is by dividing the observed polarized flux at a certain disk radius, $F_{\rm pol}(r)$, by the stellar flux incident on that disk region, $F_*/4\pi r^2$.  This number contains information on both the intrinsic albedo of particles (see e.g., \citealt{Mulders13}) and on the fraction of photons scattered toward the observer (see e.g., \citealt{Stolker16}) and is thus sometimes referred to as (polarized) geometric albedo, or contrast. This measurement is available for a relatively large number of real circumstellar disks \citep[see][]{Garufi17,Garufi18}. From our simulations, we obtained the aforementioned contrast along a {4 au-large} radial cut oriented toward the planet location. This profile is obtained from the $PI,\ Q_{\phi}$ and $U_{\phi}$ images as well, and is shown in Fig.\,\ref{fig:profile} for some illustrative cases. $Q_{\phi}$ almost always lies upon the profile of total polarized intensity, and at the planet location there is zero excess from the presence of the CPD in the case of Jupiter- and Saturn mass planets. For the higher mass planets ($\geq$ 5  $\rm{M_{Jup}}$), at the CPD location there is a contrast of {0.15 and 0.6} for $Q_{\phi}$ and $PI$, respectively. 

A locally different flux recorded in the $Q_\phi$ and $PI$ maps indicates that the pattern of the polarization diverges from centro-symmetric in the circumstellar disk (see also Sect.\,\ref{sec:CSDvsCPD}). This can be appreciated by plotting the polarization angles $\psi=0.5*\arctan(U/Q$) on top of the $Q_\phi$ maps, as done in Fig.\,\ref{fig:angles}, {zoomed to the Hill-sphere region. From the image, it is clear how the CPD changes the polarization vectors locally in the 5 and 10 M$_{\rm jup}$ planet cases. Furthermore, in the 10 M$_{\rm jup}$ case the deviation include not only the CPD ($\sim$ half of Hill-sphere), but the surrounding area as well (the spiral wakes). We plotted the signal-to-noise ratio pixel-by-pixel on Fig. \ref{fig:SNR} for the 5 Jupiter-mass case, by calculating the mean divided by the standard deviation in every pixel between the 20 RADMC-3D runs we made.

\begin{figure*}
\includegraphics[width=18cm]{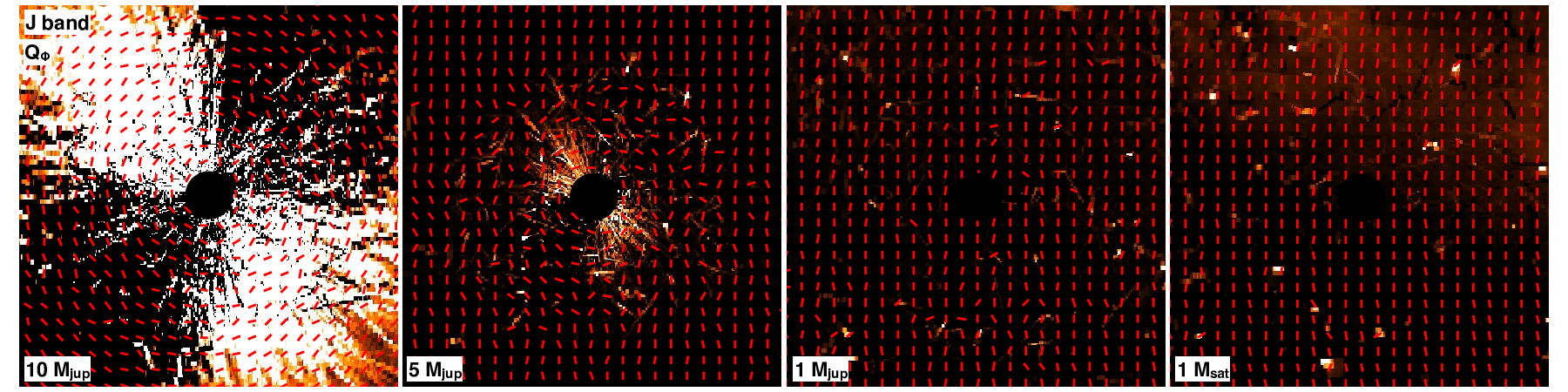}
\caption{Polarization vectors overplotted on the $Q_\phi$ maps {of the close region around the CPD. In the first two cases, a centro-symmetric pattern is visible.  In the last two cases, polarization vectors are vertical, in line with the local pattern of the protoplanetary disk}.}
\label{fig:angles}
\end{figure*}

We also extracted the contrast from the circumstellar and from the circumplanetary disk by averaging the contribution from their respective regions (0.04" for the CPD, 0.25" for the CSD that is exactly within the inner and outer edge of the disk). The values thus obtained for the circumstellar disk from the different simulations are comprised in a narrow interval of values (from 1.5\% to 3.2\%). Compared to real disks, these numbers are realistically high since the brightest disks ever observed in PDI have it up to $\sim2\%$ (see \citealt{Garufi17}). {This shows that if the CPDs have a strong enough accretion-shock surface on them, they could be surprisingly bright in some cases. }

On the other hand, the contrast obtained around the circumplanetary disk span enormously (from {4500\%} to $\lesssim0.1\%$). From the $PI$ image, the contrast of the 10 $\rm{M_{Jup}}$ case is always larger than 1 (i.e., more photons than those incident from the star are detected) indicating a strong additional source of photons to be scattered {(the hot circumplanetary disk shock surface)}. This observational scenario would be by itself a natural, robust evidence of circumplanetary disk. However, for all the other planetary mass cases that we studied the detection of the circumplanetary disk in polarized light is less straightforward. Observationally, we can define a formal threshold of 0.1\% below which the signal is mostly noise \citep{Garufi17}. According to this criterion, {7 of the remaining 9 cases (3 planet masses, 3 inclinations)} should still be regarded as detection. We caution, however, that our images does not contain extra source of noise, that could make the detections even more difficult as described here. So these detection numbers should be regarded as idealistic, best-case scenarios.

We must nonetheless consider the effect of the circumstellar disk itself that may still be present at the planet location (in particular for the 0.3 $\rm{M_{Jup}}$ case where the disk gap is more shallow than for the more massive planets) and leaves the same imprint on the scattered-light images. In this regard, we noticed that the contrast around the planet decreases toward smaller masses but then increases again for the lowest-mass case.

In Table \ref{table:contrast} we show the contrast values between the integrated Qphi and Uphi values in the CPD area, divided by the stellar flux at 50\,AU (where the planets are located from their stars). These values differ from zero, because the CPDs are marginally resolved (at 100 pc with VLT resolution power). Furthermore, where these contrast ratios are larger than $|1|$, the same applies as explained for $PI$: apart from the stellar light, the planet and the CPD shock front also contributes as photon sources.

\begin{table}
\begin{center}
\begin{tabular}{ l | c | c | c} 
 \hline
  $M_p$ [$\rm{M_{Jup}}$] & i [deg] &  $(|\sum Q_{\phi_{\rm{CPD}}}|)/F_{*}$ &  $(|\sum U_{\phi_{\rm{CPD}}}|)/F_{*}$   \\ 
  \hline
10 & 0 &        3677.6 &       163.2 \\ 
5 & 0 &        4.5 &       0.3 \\ 
1 & 0 &        1.2 &        3.8 \\ 
0.3 & 0 &       0.8 &        14.5 \\ 
10 & 30 &        5992.8 &        3232.0 \\ 
5 & 30 &        5.0 &       0.4 \\ 
1 & 30 &        1.2 &       1.3 \\ 
0.3 & 30  &      0.08 &       6.4 \\ 
10 & 60 &        6425.3 &        1761.9 \\ 
5 & 60  &       2.0 &       5.05 \\ 
1 & 60  &       2.8 &       12.3 \\ 
0.3 & 60  &       2.8 &       15.5 \\ 
 \hline
\end{tabular}
\caption{Contrast between the integrated Qphi and Uphi values in the CPD area, divided by the stellar flux intensity at 50AU. If larger than $|1|$, then not only the stellar light contributes to these values, but also the planet \& CPD shock front. Furthermore, because the values are not zeros, the CPDs are resolved with VLT at a distance of 100pc.}
\label{table:contrast}
\end{center}
\end{table}

\subsection{Circumstellar versus circumplanetary disk signal}
\label{sec:CSDvsCPD}

Our simulations show that it is formally possible to distinguish between the scattered light from the circumstellar and from the circumplanetary disk by comparing the contrast from the $PI$ and $Q_{\phi}$ images. In fact, for the two largest-mass planet scenarios the polarized contrast around the planet calculated from these two maps significantly differ {(up to a factor 50)} whereas in the 1.0 $\rm{M_{Jup}}$ case {only a minor ratio ($\sim30\%$) is visible, and no difference is appreciable in the 0.3 $\rm{M_{Jup}}$ case.} Conversely, for all our simulations the circumstellar disk signal from the $PI$ and $Q_{\phi}$ images is very similar (always within $10\%$). This behaviour can be appreciated from Fig.\,\ref{fig:ratios}. Strong discrepancies between $PI$ and $Q_{\phi}$ are expected when the scattered light deviates from a centro-symmetric pattern (see Fig.\,\ref{fig:angles}), which is the assumption under which $Q_{\phi}$ is constructed (see Eq. \ref{eq:stokesphi}). In the presence of a circumplanetary disk, photons are not expected to be scattered in such a pattern since the star is no longer the only source of photons (see on Fig. \ref{fig:scat12_incl0} for 10 Jupiter-mass case). Therefore, the comparison of the polarized contrast from the $PI$ and $Q_{\phi}$ images is a simple but potentially powerful manner to discriminate the presence of a circumplanetary disk.

\begin{figure}
\includegraphics[width=\columnwidth]{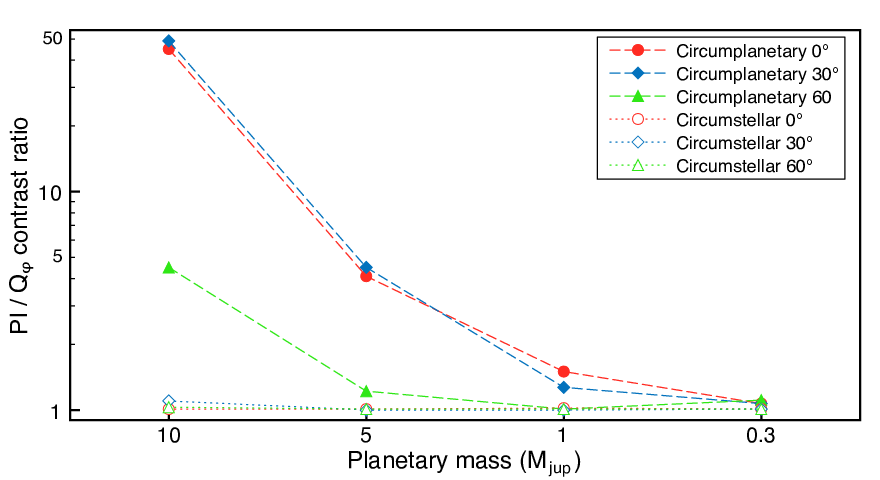}
\caption{ {$PI$/$Q_{\phi}$ contrast ratios of circumstellar- and circumplanetary disks for different planet masses and different inclinations}. From these simulations, in the 10 and 5 $\rm{M_{Jup}}$ cases, {as well as marginally in the 1 $\rm{M_{Jup}}$ case,} the signal from the circumplanetary disk can be observationally disentangled from the circumstellar disk signal.}
\label{fig:ratios}
\end{figure}

Comparing the $PI$/$Q_{\phi}$ contrast ratios of the CPD with various position angles of the planet (0, 45, 90 degrees) does not show a clear trend towards either direction (Fig. \ref{fig:PA}), hence we only show the position angle 0 on Fig. \ref{fig:ratios}. However, for all position angles considered, we found that the  $PI$/$Q_{\phi}$ contrast ratio decreases with increasing inclination angle (Fig. \ref{fig:PA}).

\section{Discussion}
\label{sec:discussion}

{The results of Sect.\,\ref{sec:results} suggest that the parallel employment of $PI$ and $Q_\phi$ maps and of multiple wavebands may help reveal a circumplanetary disk. In fact, while the light scattered off by the CPD could easily be confused for a substructure of the circumstellar disk, its polarization pattern is different. This results in a local divergence between the tangential and the total component of the polarized light, as indicated by the $Q_\phi$ and $PI$ maps respectively. In this work, we showed that this effect is appreciable for massive planets, with mass of the of at least $5\,M_{\rm jup}$.}

{As of today, the paucity of planets detected in circumstellar disks does not allow to test these predictions. Nonetheless, the ideal threshold of $5\,M_{\rm jup}$ over which the scattered light from a CPD becomes detectable is close to the common mass upper limit determined in a number of objects \citep{Claudi19,Maire17,Mesa19,Mesa19b} indicating that this approach could in principle be used in parallel to the typical differential techniques to detect the planetary thermal light.}

{The results presented here are somewhat affected by the temperature calculation as well. We used the hydrodynamic simulation calculated temperatures (via flux limited diffusion approximation), where assumed perfect thermal equilibrium between the micron sized dust and the gas (Fig. \ref{fig:temp} top-left panel). The more traditional method to calculate the dust temperature is thermal Monte Carlo computation, that can be done with RADMC-3D's \url{mctherm} command (Fig. \ref{fig:temp} top-right panel; calculated with $10^6$ photon packages). However, latter method does not include the accretional heating due to adiabatic compression, shock-heating due to accretion shock fronts, and viscous heating, all which heats up the planet vicinity. Hence, the circumplanetary region does not show up on the mctherm temperature maps (Fig. \ref{fig:temp} top-right panel), which also means that this region would not be very visible on mock observations. The difference of the two temperature calculation are shown on Fig. \ref{fig:temp} bottom panels. Everywhere else in the simulation box (i.e. in the circumstellar disk) the temperature difference between the two method is always smaller than 10-20K, the only region where the difference is at minimum 50 K (but ranges up to 1600 K difference) is the accreting planet vicinity. This test shows that for creating mock observations, the way the temperature is calculated is very important for the outcome. The shock surface of the CPD contributes to the scattering (see Fig. \ref{fig:scat12_incl0}) top row}, therefore whether this area is part of the temperature calculation (such as by using the hydrodynamic simulation temperatures) or not (with \url{mctherm}) will affect the outcome of the scattered light mock observations.

\begin{figure*}
\includegraphics[width=8.5cm]{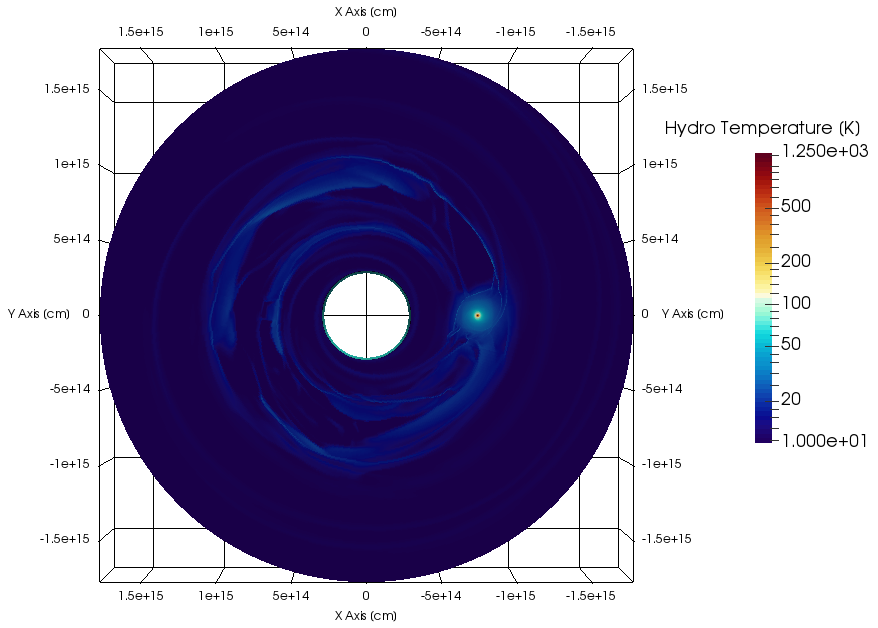}
\includegraphics[width=8.5cm]{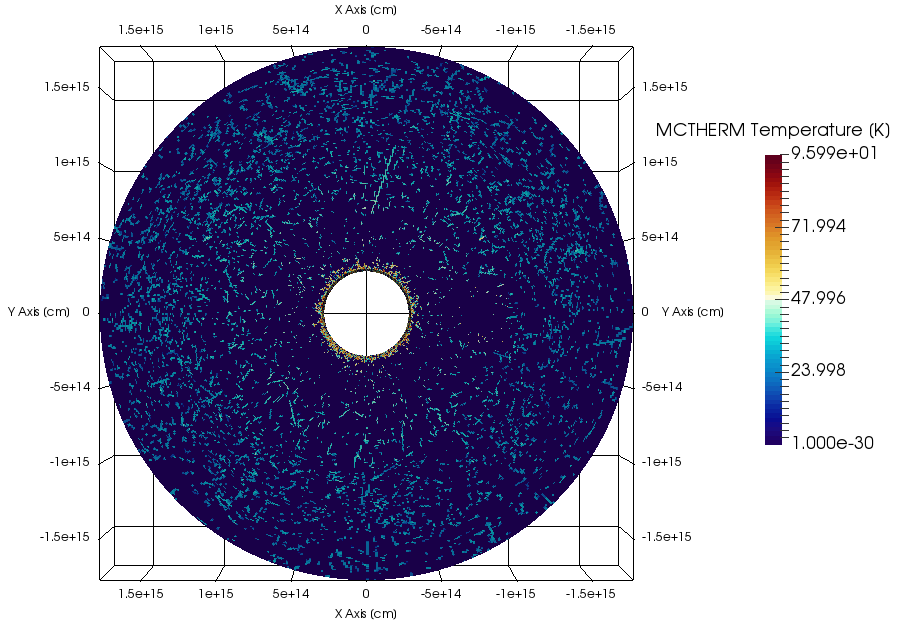}
\includegraphics[width=8.5cm]{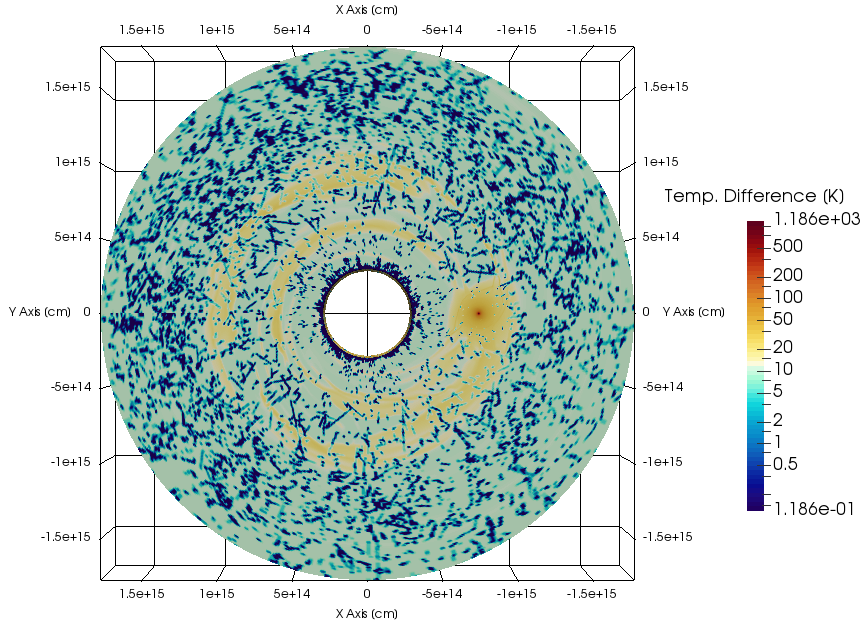}
\includegraphics[width=8.5cm]{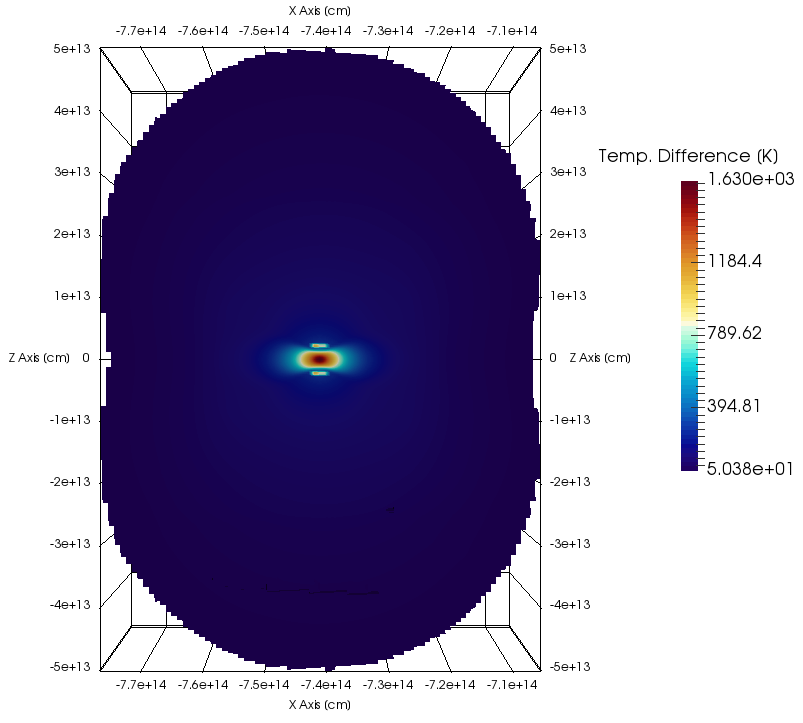}
\caption{ {Comparison of the temperature calculations in the 10 Jupiter-mass case. Top-left panel: hydrodynamic simulation calculated temperature in the midplane that include adiabatic compression, shock heating and viscous heating. Top right panel: RADMC-3D calculated temperature field in midplane with thermal Monte-Carlo (\url{mctherm}). Bottom-left panel: difference of the temperature between the hydrodynamic simulation calculated temperature and RADMC-3D's Monte-Carlo calculated one, clearly there is a large difference in the circumplanetary region. Bottom-right panel: the region in the entire simulation box, where the temperature difference between the two methods are at least 50 K -- clearly this is only the circumplanetary region (a vertical slice is shown of this area). The planet and the inner circumplanetary disk is the orange region in the middle, the shock front on the circumplanetary disk created by the accretion stream is also visible as a horizontal yellow line above the planet.}}
\label{fig:temp}
\end{figure*}

{In the 5 and 10 Jupiter-mass planet cases the circumplanetary region pops up, and some spurious photons can be seen to originate from this region (Fig. \ref{fig:scat12_incl0}). Closer inspection showed that the accretion shock front created on the surface of the circumplanetary disk (Fig. \ref{fig:shock}) due to the incoming mass influx from the circumstellar disk (via the meridional circulation; \citealt{Szulagyi14}) is the origin of the noisiness on the mock images on Fig. \ref{fig:scat12_incl0}. This shock front is strong and hot ($>1000$\,K) enough only in the 5 and 10 Jupiter-mass cases, not for the smaller planets. This is due to the the fact that the incoming accretion flow velocity scales with the planetary mass (nearly free-fall velocity). The higher velocity of the influx creates stronger and hotter shock fronts on the circumplanetary disk surface, with increasing planetary mass \citep{Szulagyi17gap}. This shock-front is helping the observability of the circumplanetary disk, hence it is also important to self-consistently include it when creating mock observations on circumplanetary disks.}

\begin{figure}
\includegraphics[width=\columnwidth]{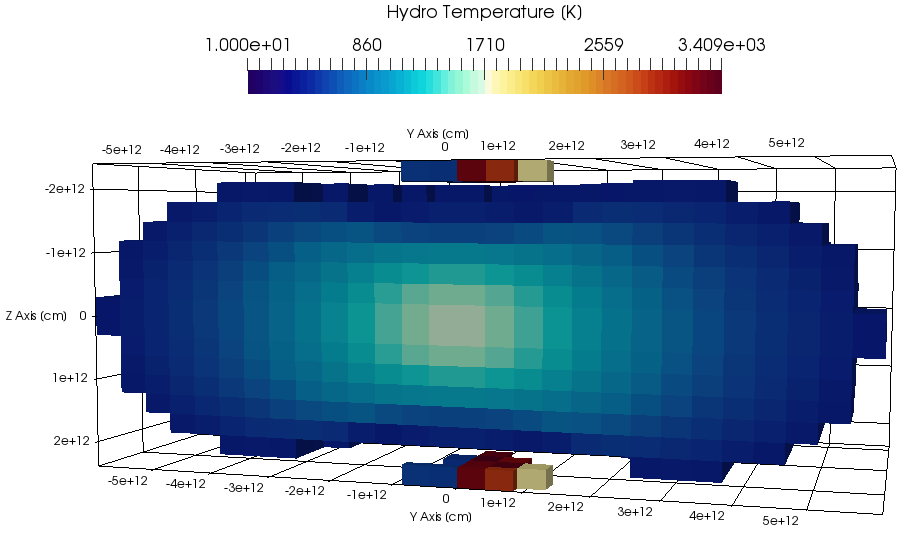}
\caption{ {Hydrodynamic simulation calculated temperature field (threshold) in the circumplanetary disk (10 Jupiter-mass case), where the accretion shock-front ($>1000$\,K) on the disk surface is visible by yellow, orange and red areas above the planet. The noisiness of the circumplanetary region in the mock observations (e.g. Fig. \ref{fig:scat12_incl0}) is due to this shock front. However, this shock front is also helping the detection of the circumplanetary disk.}}
\label{fig:shock}
\end{figure}

The results also depend on the optical depth. In this work we took care of dust evaporation above silicate evaporation temperature (1500 K), which meant that the dust density were put to zero, where the temperatures were rising above this limit (this region is near the planet, e.g. in the shock front on the circumplanetary disk surface). The optically thick regions near the planet could absorb some of the emitted photons, that will reduce the irradiation in the outer CPD. This effect will also change the the scattering, therefore the polarized images.

Our models only covers part of the parameter space. We assumed a fixed dust-to-gas ratio of 0.01 even in the circumplanetary disk \citep{DSz18}, however real disks can have smaller and larger values than this \citep[e.g.][]{YG05,Marel13,DD14,Birnstiel12,WB14,Ansdell16}, which might affect the results. In this work we have considered the planets to be 50 AU from the star, but the circumplanetary disk-circumstellar disk contrast can be very different if the circumplanetary disk at another distance. Circumplanetary disks closer to the star tend to be more optically thick, and hotter than the more distant ones.

For the circumstellar disk mass we considered an average value of 0.01 $\rm{M_{Sun}}$, and the radial extent was between 20-120 AU, similar to a transitional disk with an inner cavity. While the circumplanetary disk mass linearly scales with the circumstellar disk mass \citep{Szulagyi17gap}, the changes in mass will also result in different optical depth, which can affect the results described here. The large, optically thin inner 20 AU can also affect the results.

The hydrodynamic simulations did not include magnetic fields, e.g. the fields of the disks, which might affect the dust density distribution \citep{Gressel13}.  
Since the planet interior structure was not part of the simulation, any temperature in the planet region is a lower limit to the radiation from the planet. This will affect the radiation field in/from the CPD, as well as its spatial structure.


\section{Summary}

In this work we investigated polarized scattered light detectability of circumplanetary disks surrounding nascent planets. We ran hydrodynamic simulations with mesh refinement to resolve sufficiently the circumplanetary disk. We used radiative transfer included hydrodynamics, to realistically estimate the temperature. Then, we post-processed the simulations with RADMC-3D Monte-Carlo radiative transfer software to create polarized light images in J band. We added convolution with a PSF-size at the diffraction limit, assuming 8.2 meter mirror, like VLT. 

We considered different planetary mass cases: Saturn, 1, 5, 10 $\rm{M_{Jup}}$, different disk inclinations of 0, 30, 60, 90 degrees, { and various planetary locations (0, 45, 90 degrees).} The planets were embedded in a  0.01 $\rm{M_{Sun}}$ circumstellar disk, 50 AU away from their star, which was assumed to be a Sun-equivalent. 

Our $I,\ PI,\ Q_{\phi}$ and $U_{\phi}$ images revealed that the circumplanetary disk detection is only possible in the case of very massive planets (5 and 10 $\rm{M_{Jup}}$), although it is highly dependent on how optically thick is the circumplanetary disk (i.e. how much dust it contains, and what is the temperature there, { whether the dust is evaporated or not). In these high mass planet cases the accretion shock front on the surface of the circumplanetary disk \citep{SzM17} is so strong and luminous, that it helps the observability of this subdisk. Therefore the inclusion of this shock-front is important when modelling observability of the circumplanetary disk.} 

The circumplanetary disk detection is challenging in polarized light, not only because of sensitivity but also due to the contrast with the circumstellar disk. However, we showed that, ideally speaking, it is possible to distinguish between the two disk's contributions by comparing the total polarized light (from the $PI$ image) and the centro-symmetric polarized light (from the $Q_\phi$ image), as well as by finding stronger polarized colors in the circumplanetary disk than in the neighboring circumstellar disk.

In conclusion, while circumplanetary disk detection might be challenging in polarized light, the  $PI$/$Q_\phi$ images can be possible tools to detect the circumplanetary disk within the circumstellar disk.

\section*{Acknowledgments}

We thank for Adriana Pohl providing the opacity table, including the polarization matrix. J.Sz. thanks for the financial support through the Swiss National Science Foundation (SNSF) Ambizione grant PZ00P2\_174115. Furthermore, these results are part of a project that has received funding from the European Research Council (ERC) under the European Union’s Horizon 2020 research and innovation programme (Grant agreement No. 948467). We also acknowledge support from the project PRIN-INAF 2016 The Cradle of Life - GENESIS-SKA (General Conditions in Early Planetary Systems for the rise of life with SKA) and  from INAF/Frontiera (Fostering high ResolutiON Technology and Innovation for Exoplanets and Research in Astrophysics) through the "Progetti Premiali" funding scheme of the Italian Ministry of Education, University, and Research. Computations partially have been done on the "Piz Daint" machine hosted at the Swiss National Computational Centre and partially carried out on ETH Z\"urich's Euler machine. 

\section*{Data Availability}

The data underlying this article will be shared on reasonable
request to the corresponding author.

\begin{appendix} 
\section{Additional Figures} \label{All_maps}
{Similarly to Fig.\,\ref{fig:scat12_incl0}, Figs.\,\ref{fig:scat12_incl30}, \ref{fig:scat12_incl60} show the $I,\ PI,\ Q_{\phi}$ and $U_{\phi}$ maps for the four planetary masses explored. In particular, Fig.\,\ref{fig:scat12_incl30} shows 30 deg inclination and Fig. \ref{fig:scat12_incl60} represents 60 deg inclination.}

Fig. \ref{fig:SNR} shows the signal-to-noise ratio in every pixel for Fig. \ref{fig:angles} 5 Jupiter-mass case, in order to get an idea about the accuracy of the individual pixel fluxes.

Fig. \ref{fig:PA} represents the $PI/Q_{\phi}$ contrast ratios for the CPDs with various inclination, and various position angle of the planet.

\begin{figure*}
\includegraphics[width=18cm]{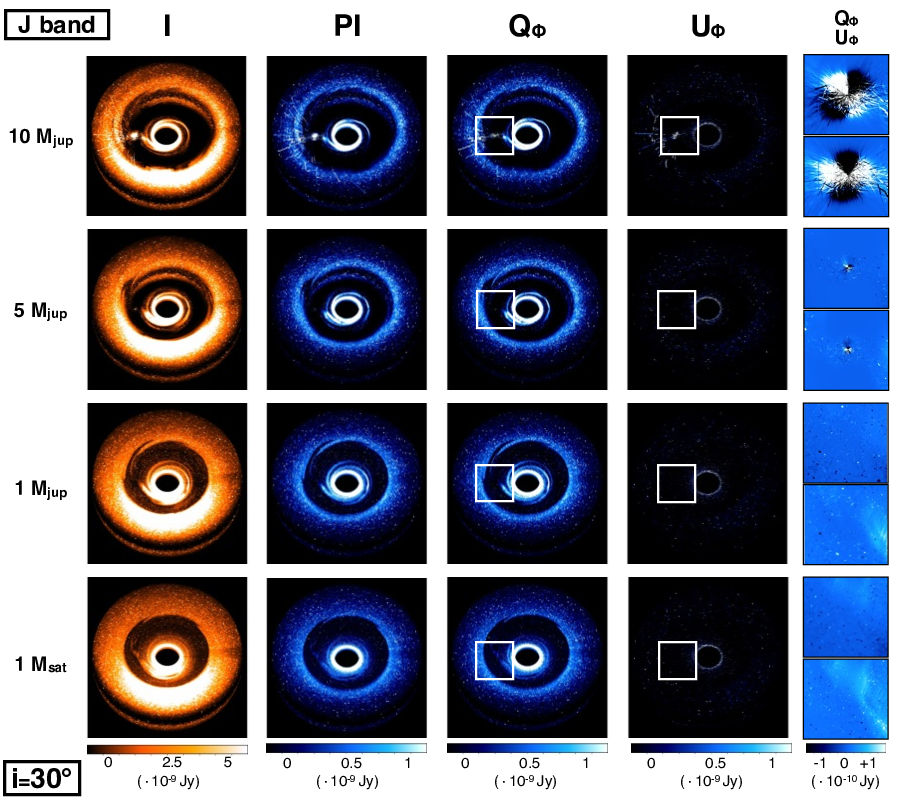}
\caption{Same as Fig. \ref{fig:scat12_incl0}, but for 30$^\circ$ inclination.}
\label{fig:scat12_incl30}
\end{figure*}

\begin{figure*}
\includegraphics[width=18cm]{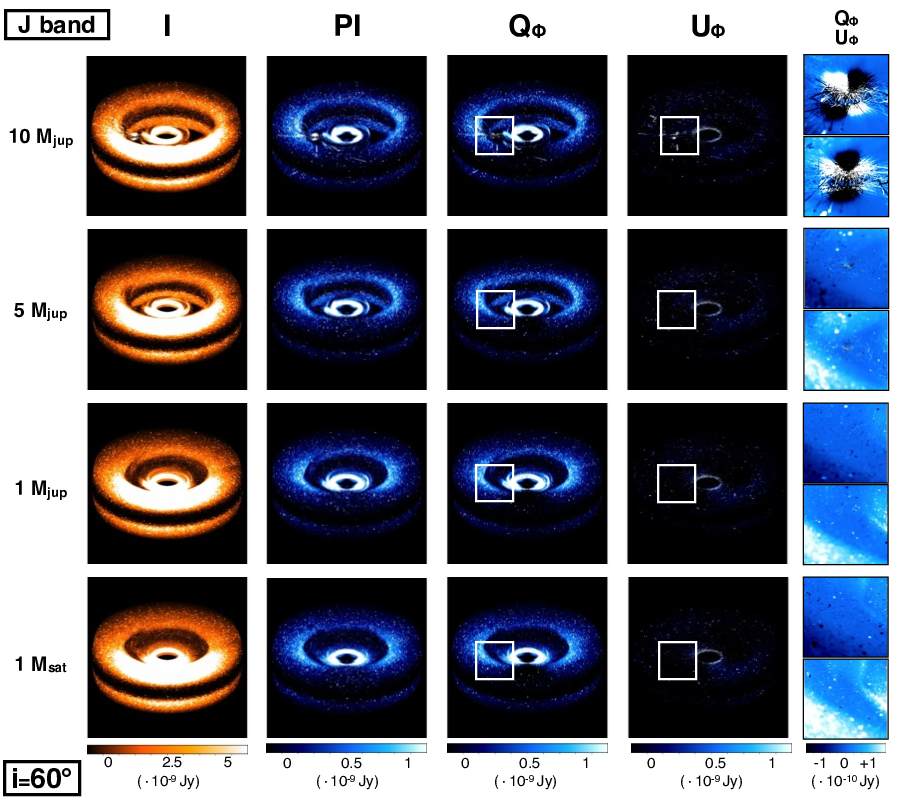}
\caption{Same as Fig. \ref{fig:scat12_incl0}, but for 60$^\circ$ inclination.}
\label{fig:scat12_incl60}
\end{figure*}

\begin{figure}
\includegraphics[width=\columnwidth]{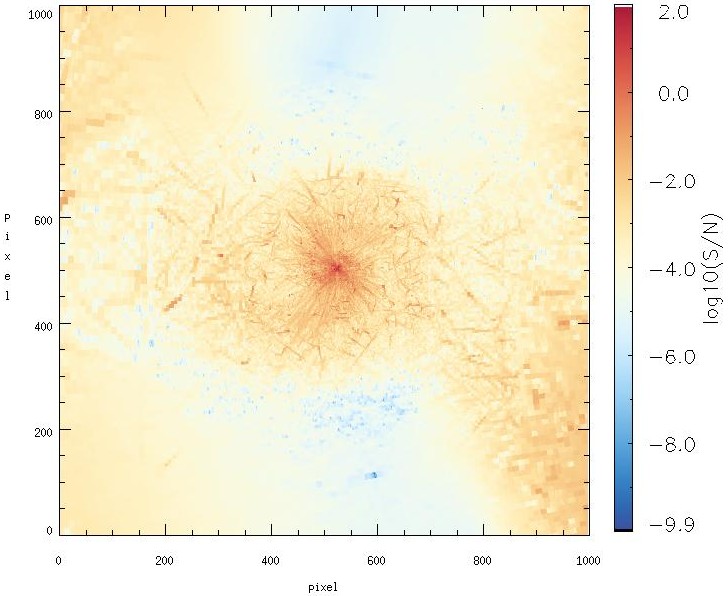}
\caption{Signal-to-noise ratio in every pixel for Fig. \ref{fig:angles} second panel (5 Jupiter-mass case, zoom to the circumplanetary region).}
\label{fig:SNR}
\end{figure}

\begin{figure}
\includegraphics[width=\columnwidth]{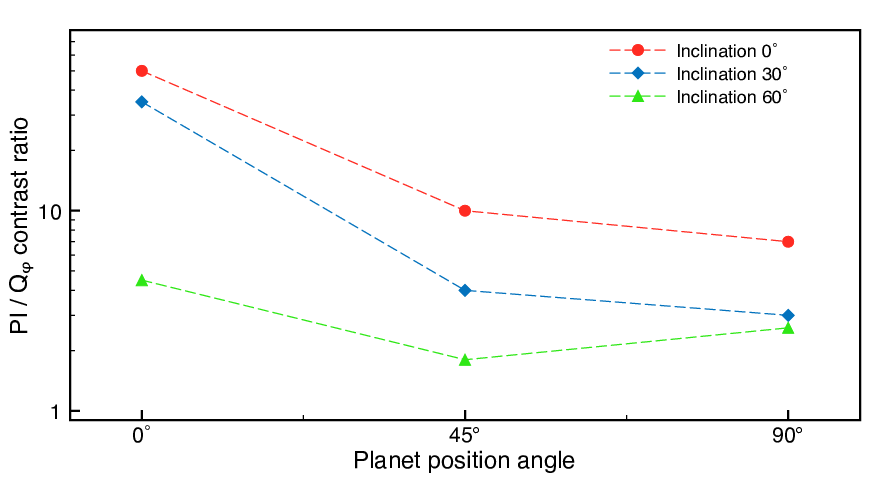}
\caption{Contrast ratio of $PI/Q_{\phi}$ for the CPD depending on the planet position angle and the inclination. }
\label{fig:PA}
\end{figure}

\end{appendix}

\label{lastpage}

\end{document}